\documentclass[conference]{IEEEtran}

\usepackage[ruled,vlined,linesnumbered]{algorithm2e}
\usepackage{amsmath,amssymb,amsfonts}
\usepackage[english]{babel}
\usepackage{balance}
\usepackage[scaled]{beramono}
\usepackage{booktabs}
\usepackage{cite}
\usepackage[T1]{fontenc}
\usepackage{graphicx}
\usepackage[utf8]{inputenc}
\usepackage{listings}
\usepackage{microtype}
\usepackage{nameref}
\usepackage{newtxtext,newtxmath}
\usepackage[group-separator={,}]{siunitx}
\usepackage[subrefformat=parens]{subcaption}
\usepackage[normalem]{ulem}
\usepackage{url}
\usepackage{xcolor}

\graphicspath{{figs}}

\definecolor{color-bg}{HTML}{F6F8FA}
\definecolor{color-keyword}{HTML}{D73A49}
\definecolor{color-ident}{HTML}{005CC5}
\definecolor{color-string}{HTML}{032F62}
\definecolor{color-comment}{HTML}{6A737D}

\usepackage{tikz}

\newcommand\copyrighttext{%
  \footnotesize \textcopyright \the\year{} IEEE. Personal use of this material is permitted.  Permission from IEEE must be obtained for all other uses, in any current or future media, including reprinting/republishing this material for advertising or promotional purposes, creating new collective works, for resale or redistribution to servers or lists, or reuse of any copyrighted component of this work in other works.}

\newcommand\copyrightnotice{%
\begin{tikzpicture}[remember picture,overlay]
\node[anchor=south,yshift=10pt] at (current page.south) {\fbox{\parbox{\dimexpr0.75\textwidth-\fboxsep-\fboxrule\relax}{\copyrighttext}}};
\end{tikzpicture}%
}

\begin{document}

\title{Modernizing an Operational Real-time Tsunami Simulator to Support Diverse Hardware Platforms}

\makeatletter
\newcommand{\linebreakand}{%
  \end{@IEEEauthorhalign}
  \hfill\mbox{}\par
  \mbox{}\hfill\begin{@IEEEauthorhalign}
}
\makeatother

\author{
    \IEEEauthorblockN{Keichi Takahashi}
    \IEEEauthorblockA{Tohoku University \\
    Sendai, Japan \\
    keichi@tohoku.ac.jp}
    \and
    \IEEEauthorblockN{Takashi Abe}
    \IEEEauthorblockA{Tohoku University\\
    Sendai, Japan \\
    t-abe@irides.tohoku.ac.jp}
    \and
    \IEEEauthorblockN{Akihiro Musa}
    \IEEEauthorblockA{NEC Corporation \\
    Tokyo, Japan \\
    a-musa@nec.com}
    \linebreakand
    \IEEEauthorblockN{Yoshihiko Sato}
    \IEEEauthorblockA{NEC Solution Innovators \\
    Tokyo, Japan \\
    yoshihiko.sato@nec.com}
    \and
    \IEEEauthorblockN{Yoichi Shimomura}
    \IEEEauthorblockA{NEC Solution Innovators \\
    Tokyo, Japan \\
    shimomura32@nec.com}
    \and
    \IEEEauthorblockN{Hiroyuki Takizawa}
    \IEEEauthorblockA{Tohoku University\\
    Sendai, Japan \\
    takizawa@tohoku.ac.jp}
    \and
    \IEEEauthorblockN{Shunichi Koshimura}
    \IEEEauthorblockA{Tohoku University\\
    Sendai, Japan \\
    koshimura@irides.tohoku.ac.jp}
}

\maketitle
\copyrightnotice

\begin{abstract}
    To issue early warnings and rapidly initiate disaster responses after tsunami damage, various
    tsunami inundation forecast systems have been deployed worldwide. Japan’s Cabinet Office
    operates a forecast system that utilizes supercomputers to perform tsunami propagation and
    inundation simulation in real time. Although this real-time approach is able to produce
    significantly more accurate forecasts than the conventional database-driven approach, its wider
    adoption was hindered because it was specifically developed for vector
    supercomputers. In this paper, we migrate the simulation code to modern CPUs and GPUs in a
    minimally invasive manner to reduce the testing and maintenance costs. A directive-based
    approach is employed to retain the structure of the original code while achieving performance
    portability, and hardware-specific optimizations including load balance improvement for GPUs are
    applied. The migrated code runs efficiently on recent CPUs, GPUs and vector processors: a
    six-hour tsunami simulation using over 47 million cells completes in less than 2.5 minutes on 32
    Intel Sapphire Rapids CPUs and 1.5 minutes on 32 NVIDIA H100 GPUs. These results demonstrate
    that the code enables broader access to accurate tsunami inundation forecasts.
\end{abstract}


\section{Introduction}

Throughout the history of humanity, tsunamis have been one of the deadliest natural disasters. The
United Nations Office for Disaster Risk Reduction (UNDRR) reported in 2018 that between 1998 and
2017, tsunamis caused \num{251770} human casualties and US\$280 billion economic
losses\footnote{\url{http://www.undrr.org/quick/10744}}. To reduce the damage inflicted by
tsunamis, various tsunami warning systems have been developed and deployed since the mid-20th
century~\cite{Bernard2015}.

The approaches for tsunami inundation forecast are roughly categorized into two: (1)~\emph{database-driven
approach} and (2)~\emph{real-time forward approach}. In the database-driven approach, a large number of
tsunami scenarios are simulated in advance and their outcomes are stored in a database. Once an
earthquake occurs, this database is queried to find the closest tsunami scenario. Although this
approach is advantageous in terms of speed, it is fundamentally limited by the coverage of the
tsunami scenario database and cannot offer accurate estimates for unanticipated earthquake
scenarios. The real-time forward approach, on the other hand, starts a tsunami propagation and
inundation simulation immediately after an earthquake occurs. This allows accurate simulation by incorporating
the latest information such as topography, bathymetry, tidal conditions and coastal protection
facilities. In the database-driven approach, the entire database needs to be rebuilt every time such
simulation conditions change. The primary downside of the real-time forward approach is its
computational cost, which usually requires the use of High-Performance Computing (HPC) systems.

We have developed the world's first operational tsunami inundation forecast system based on the
real-time forward approach~\cite{Inoue2019,Musa2019,Musa2018}.
This system achieves the so-called \emph{10-10-10 (triple ten)
challenge}: it estimates the tsunami fault model in 10 minutes and subsequently completes a tsunami
inundation and damage simulation in 10 minutes with a \qty{10}{m}-grid resolution. This is made
possible by harnessing the computing power of NEC's vector supercomputers such as
SX-ACE~\cite{Egawa2017} and SX-Aurora TSUBASA~\cite{Komatsu2018,Egawa2020,Takahashi2023}. These
supercomputers are well-known for their world-class memory access performance, which is vital for
accelerating generally memory-bound tsunami simulations. The tsunami inundation forecast system is
approved by the Meteorological Agency of Japan and provides forecasts to a few local governments
including Kochi Prefecture. Furthermore, this forecast system has been integrated into the
nationwide Disaster Information System (DIS) operated by the Cabinet Office of Japan since 2018. The
nationwide system is deployed at computing centers at Tohoku University and Osaka University in
Japan to enable geo-redundancy.

Although our system could potentially be spread out to local governments across Japan and even
worldwide to help tackle tsunami disasters, widespread adoption has been challenging mainly due to
its limited execution platforms. At the time when SX-ACE and SX-Aurora TSUBASA were released, they
offered the world-highest memory bandwidth. It was thus a natural and optimal choice to target these
systems. However, with the advent of High-Bandwidth Memory (HBM)~\cite{Jun2017}, modern GPUs and
CPUs now offer memory bandwidth in the order of TB/s, comparable to that of the SX systems. These
processors are thus promising alternatives for running the tsunami inundation simulation code.

In this paper, we aim to migrate our tsunami propagation and inundation simulation code to CPUs and
GPUs. Since the code is already deployed and used in a production system, a complete rewrite or
significant refactoring would require extensive testing and verification as well as rework on the
existing system. Moreover, implementing a variation for each target processor is infeasible due to
high development and maintenance costs. Our unique challenge is thus to modernize the original code
in a minimally invasive way into a performance-portable version that can run across multiple
hardware platforms.

The contributions of this paper are summarized as follows:

\begin{itemize}
    \item We migrate an operational, production-scale tsunami simulation code employed in a
        government disaster information system. Unlike most performance portability studies that
        rewrite or refactor large portions of the code, we aim to minimize source code
        modifications.
    \item We migrate a code specifically designed for long-vector architectures to recent GPUs and
        CPUs. These architectures significantly differ in various aspects such as the hierarchy of
        parallelism and degree of parallelism at each level. Combined with the irregular loop
        structure inherent to the code, this poses a significant challenge.
    \item We evaluate the performance of the migrated tsunami simulation code on multiple
        supercomputers equipped with recent CPUs, GPUs and vector processors. We conduct an
        extensive evaluation to quantify the benefit of our target-specific but non-intrusive
        performance optimizations.
\end{itemize}

The rest of this paper is structured as follows. Section~\ref{sec:background} briefly describes the
design of the original tsunami simulation code. Section~\ref{sec:related-work} reviews recent
studies on tsunami simulation. Section~\ref{sec:proposal} presents the migration process of the
tsunami real-time simulator to CPUs and GPUs. Section~\ref{sec:evaluation} evaluates our simulator
on multiple HPC systems. Section~\ref{sec:conclusion} concludes this paper.

\section{Real-time Tsunami inundation  model}\label{sec:background}

In this section, we describe the basic design and implementation of the original tsunami simulation code
named the \emph{Real-time Tsunami inundation (RTi)} model.
We first outline the underlying numerical model and then describe its implementation and
parallelization for NEC SX-series supercomputers.

\subsection{Numerical model}

The RTi model is based on the Tohoku University Numerical
Analysis Model for Investigation of tsunamis (TUNAMI)~\cite{Goto1997,Imamura2006}, which is a
well-established tsunami model widely known in the research community and approved by the United
Nations Educational, Scientific and Cultural Organization (UNESCO). TUNAMI encompasses several
variations using different governing equations and coordinate systems. The RTi model uses the
TUNAMI-N2 model, which specializes in near-field tsunamis and solves the two-dimensional non-linear
shallow water equations in a Cartesian coordinate. In the following, the numerical model is briefly
summarized. The equation of continuity is as follows:
\begin{align}
    \frac{\partial \eta}{\partial t}+\frac{\partial M}{\partial x}+\frac{\partial N}{\partial y}&=0, \label{eqn:eq1}
\end{align}
where $\eta$ is the vertical displacement of the water surface, i.e., water level, and $M$ and $N$
are the discharge fluxes in the x- and y-directions, respectively. The equation of motion is:
\begin{align}
    \frac{\partial M}{\partial t}+\frac{\partial}{\partial x}\left(\frac{M^2}{D}\right)
        +\frac{\partial}{\partial y}\left(\frac{M N}{D}\right)
        +g D \frac{\partial \eta}{\partial x} \nonumber \\
        +\frac{g n^2}{D^{7 / 3}} M \sqrt{M^2+N^2}&=0, \label{eqn:eq2} \\
    \frac{\partial N}{\partial t}+\frac{\partial}{\partial x}\left(\frac{M N}{D}\right)
            +\frac{\partial}{\partial y}\left(\frac{N^2}{D}\right)
            +g D \frac{\partial \eta}{\partial y}\nonumber \\
            +\frac{g n^2}{D^{7 / 3}} N \sqrt{M^2+N^2}&=0, \label{eqn:eq3}
\end{align}
where $D$ is the total water depth, $g$ is the gravity constant and $n$ is Manning's roughness
coefficient. Note that $D=h+\eta$ where $h$ is the still water depth.

Equations~(\ref{eqn:eq1}) to (\ref{eqn:eq3}) are solved using a leap-frog time-stepping scheme and
staggered finite-difference grid. The code employs a nested grid system, in which multiple grids
with different spatial resolutions are combined. The main reason for using a nested grid system is
to cope with the multi-scale nature of tsunamis. Technically, the following Courant-Friedrichs-Lewy
(CFL) condition must be met for numerical stability:
\begin{equation}
    \frac{\Delta x}{\Delta t} \geq \sqrt{2gh_{\mathrm{max}}},
\end{equation}
where $h_{\mathrm{max}}$ is the maximum still water depth. To keep $\Delta t$ constant, one can use
a smaller $\Delta x$ in coastal regions and a larger $\Delta x$ in deep waters. In this paper, we
use a refinement ratio of 3:1, i.e., the ratio between the spatial resolutions of the parent and
child grid is 3:1. The nesting is inclusive, meaning that a child grid is always fully enclosed by
its parent grid. Each grid level is composed of multiple rectangular grids (hereinafter referred to
as \emph{blocks}).

Figure~\ref{fig:nesting} illustrates an example of a nested grid with three nest levels around
Kochi Prefecture in Japan that faces the Pacific Ocean. In this illustration, darker shades
of blue indicate deeper waters. It can be seen that the finer grid levels track the coastlines and
avoid deeper water.

\begin{figure}
\centering
\includegraphics[width=0.9\linewidth]{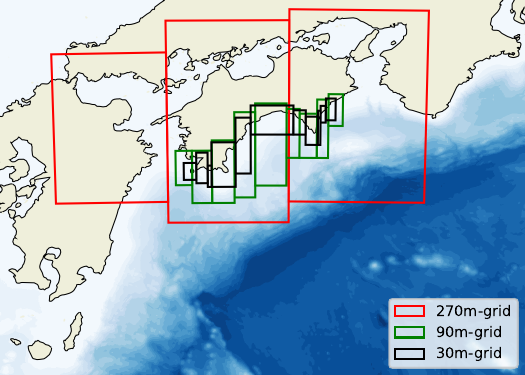}%
\caption{An example of three nested grid levels in the coast of Kochi Prefecture.}\label{fig:nesting}
\end{figure}

\subsection{Implementation}

The RTi model was initially developed for the NEC SX-ACE system~\cite{Musa2018,Inoue2019}, and
later ported to the NEC SX-Aurora TSUBASA system~\cite{Musa2019} equipped with NEC Vector Engine
(VE) processors~\cite{Komatsu2018,Egawa2020,Takahashi2023}. The original code is implemented in
Fortran 90 and parallelized using MPI only\footnote{We evaluated a work-in-progress
hybrid MPI/OpenMP version, but it generally performed poorer than the flat MPI version. It is also
not used in the operational tsunami forecast system.}. NEC's Fortran compiler is used to vectorize
the code with the help of some vendor compiler directives to facilitate vectorization. Since neither
intrinsics nor inline assembly is used, the code is functionally portable, but not necessarily
performance-portable.

The domain decomposition is static and supplied via a configuration file by the user. One or more MPI ranks are
assigned to each grid level. Each block in a grid level can be further decomposed into multiple ranks.
In this case, one-dimensional domain decomposition is used. Although two-dimensional
decomposition is preferable in terms of communication volume, it shortens the vectorized innermost
loop. Since the vector register of a VE is \qty{16384}{\bit}-wide and thus requires a long loop
length for efficiency, one-dimensional decomposition is chosen over two-dimensional decomposition. Each rank is
always assigned consecutive blocks. Further implementation details of the original code are
described by its developers in their previous studies~\cite{Musa2019,Inoue2019}.

\section{Related work}\label{sec:related-work}

Tsunami-HySEA~\cite{Macias2016,Macias2017,Macias2020} is a state-of-the-art tsunami simulation code
being developed at the University of M\'{a}laga that aims at realizing Faster Than Real Time (FTRT)
tsunami simulation by utilizing GPUs. Tsunami-HySEA is developed in CUDA and uses
MPI for multi-GPU parallelization. Similar to the code in this paper, Tsunami-HySEA also employs a
nested grid system~\cite{Macias2016,Macias2017} to reduce the computational load. The authors
reported that an \qty{8}{h}-simulation of a tsunami hitting the Gulf of C\'{a}diz with
5.5 million cells took 2.5 minutes to complete~\cite{Gaite2022}. The code is highly tuned for
NVIDIA GPUs, and takes advantage of features such as GPUDirect RDMA for inter-GPU communication,
asynchronous CPU-GPU memory transfer and asynchronous file I/O.

TRITON-G~\cite{Acuna2018} is another GPU-accelerated tsunami simulation code. The code is also
developed in CUDA, and solves the non-linear spherical shallow water equations. Unlike the
common nested grid system approach, this code adopts a quadtree-based approach for mesh refinement
and Hilbert space-filling curve for load balancing. TRITON-G completed a \qty{10}{\hour}-simulation
covering a large portion of the Indian Ocean with approximately 33 million cells in under
\qty{10}{\minute} using three NVIDIA Tesla P100 GPUs. This code was deployed as an operational
model at the Regional Integrated Multi-Hazard Early Warning System for Africa and Asia (RIMES).
In addition to Tsunami-HySEA and TRITON-G, a number of GPU-accelerated tsunami simulation codes have
been developed in previous studies, most of which are implemented either in
CUDA~\cite{Koshimura2010,Satria2012} or CUDA Fortran~\cite{Yuan2020}.

JAGURS~\cite{Baba2014,Baba2016} is a tsunami code originally based on the URSGA~\cite{Jakeman2010}
hydrodynamic modeling tool. JAGURS significantly enhanced the scalability of URSGA by
introducing a nested grid system and parallelizing the code using MPI and OpenMP. The performance of
JAGURS was tested on the K computer, Japan's former flagship supercomputer~\cite{Baba2016}.
JAGURS was able to complete a \qty{20}{s}-simulation of a large-scale, high-resolution model
consisting of 100 billion cells in \qty{82}{s} using the entire \num{82844} nodes of the K computer.
The model covered an area of \qty{1000}{km} $\times$ \qty{780}{km} and contained all of southwestern
Japan. In addition, the authors also executed a \qty{5}{h}-simulation of a tsunami in the Nankai
region of Japan with 680 million cells in \qty{6625}{s} using \num{8748} nodes.

EasyWave~\cite{Christgau2014,Christgau2020} is a simple tsunami simulation code that solves the
linear long-wave theory equations in spherical coordinates. Since the nonlinear bottom friction is
ignored, it is mainly intended for modeling far-field tsunamis. The code was originally developed as
a sequential code and later ported to OpenMP, CUDA, OpenACC~\cite{Christgau2014} and
SYCL~\cite{Christgau2020}. EasyWave does not employ MPI parallelism and thus does not scale out to
multiple nodes, making it unsuitable for real-time tsunami simulation.

Although various tsunami simulation codes have been developed in the past, they are either optimized
for a specific hardware platform ~\cite{Gaite2022,Acuna2018,Koshimura2010,Satria2012,Yuan2020}
or small-scale and simplified for performance
analysis~\cite{Christgau2020,Buttner2024}. To the best of our knowledge, this paper is the first
work to present the migration process of a production-scale real-time tsunami simulation code
to multiple hardware platforms.

\section{Modernizing an operational tsunami simulator to support diverse hardware platforms}\label{sec:proposal}

In this section, we first analyze the original code and then present our migration approach to
support modern CPUs and GPUs. We further enhance the scalability of the migrated code by optimizing
communication performance and improving load balance.

\subsection{Analysis of the original RTi model}

Figure~\ref{fig:subroutines} summarizes the subroutines called in the time integration loop. The
time integration loop starts by solving Equation~(\ref{eqn:eq1}), which calculates the water level
$\eta$ with the discharge fluxes $M$ and $N$. The obtained $\eta$ is sent from the child grid to the
parent grid, and the boundary cells are exchanged between neighboring ranks in the same grid level.
Then, Equations~(\ref{eqn:eq2}) and (\ref{eqn:eq3}) are solved, which calculate the discharge
fluxes, $M$ and $N$. Subsequently, the discharge fluxes are sent from the parent to the child grid,
and the boundary cells are exchanged between neighbors. An iteration finishes by updating output
data such as maximum water velocities, water levels and inundation depths, and swapping the double
buffers. Out of the routines in Fig.~\ref{fig:subroutines}, NLMASS (Equation~\ref{eqn:eq1}) and
NLMNT2 (Equations~\ref{eqn:eq2},~\ref{eqn:eq3}) account for 60--70\% of the runtime on an SX-Aurora
TSUBASA system. Outside the time integration loop, there exist a number of routines including file
I/O, initialization of data structures, fault model estimation and damage estimation. However, the
total runtime of these routines is negligible compared to that of the time integration loop, and
thus they are not the scope of this paper.

\begin{figure}
\centering
\includegraphics[scale=0.9]{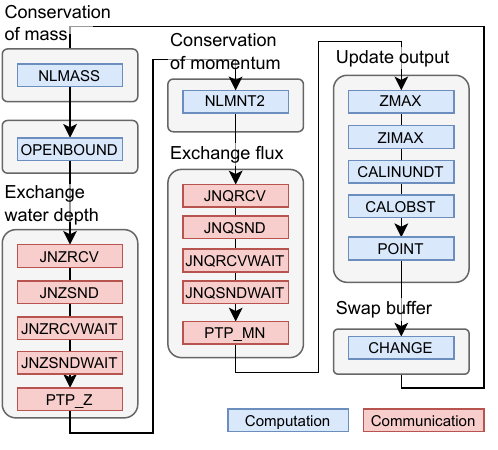}
\caption{Overview of subroutines in the time integration loop.}\label{fig:subroutines}
\end{figure}

Because of the employed finite-difference scheme, major computational kernels are two-dimensional
stencil loops where each cell in a block can be updated independently. One key feature of this code
is that a grid level is composed of many blocks of different sizes, and thus a loop that iterates
over the blocks in a grid level exists.

Listing~\ref{lst:loop-structure} shows the high-level structure of major loops, including the loop nests in the
NLMASS and NLMNT2 routines. The outermost KK loop iterates over the blocks in a grid level. The
inner I and J loops scan every cell in a single block. Although the code is designed in a
configurable way such that the I and J loops can be either mapped to the y- or x-directions, in
this paper we assume that the I and J loops iterate over the y- and x-directions of a block,
respectively. There are no dependencies between the KK, J and I loops, and they can thus be executed in
any order. There may be multiple J-I loop nests with different trip counts in a single KK loop.

\begin{lstlisting}[caption={Basic structure of major loops.},label={lst:loop-structure},float]
DO KK = BLK_PROC_ST, BLK_PROC_ED
  ! ...
  DO J = JST_MN_X2(KK), JED_MN_X2(KK)
    DO I = IST1_OWN(KK), IED2_OWN(KK)
      ! Update cell
    END DO
  END DO
  ! ...
END DO
\end{lstlisting}

It should be noted that the trip counts of the I and J loops depend on the loop index
of the outer KK loop. This irregular loop structure cannot be simply collapsed using the
\verb|collapse| clause available in OpenACC and OpenMP, because the trip count of every nested loop to
be collapsed needs to be predetermined and invariant. Furthermore, the trip count of the KK loops is
short. For example, our dataset contains 60 blocks in total, which are distributed over multiple
ranks. Thus, the KK loop alone does not provide enough parallelism for both GPUs and CPUs. The trip
counts of both the J and I loops are in the order of \num{10000}.

\begin{lstlisting}[caption={Basic structure of major loops after migration.},label={lst:offload-loop},float]
!$omp parallel
DO KK = BLK_PROC_ST, BLK_PROC_ED
  ! ...
  !$acc kernels async(MOD(KK,NSTREAMS))
  !$acc loop collapse(2)
  !$omp do
  DO J = JST_MN_X2(KK), JED_MN_X2(KK)
    !$omp simd
    DO I = IST1_OWN(KK), IED2_OWN(KK)
      ! Update cell
    END DO
    !$omp end simd
  END DO
  !$omp end do nowait
  !$acc end kernels
  ! ...
END DO
!$omp end parallel
!$acc wait
\end{lstlisting}

\subsection{Migration of bottleneck routines}\label{sec:basic-design}

As described earlier, the code is already integrated into multiple tsunami early warning systems
 for providing tsunami inundation forecasts.
Thus, rewriting a large portion of the codebase or porting the entire codebase to a different
programming language or programming model is infeasible. We therefore aim to take a strategy that
minimizes the code modifications.

The programming models we have considered are CUDA Fortran, OpenACC, OpenMP Target Offloading
and Fortran Do Concurrent (DC). CUDA Fortran is a kernel-based programming model and requires
considerable restructuring of the code; hence it is unsuitable for this migration. Fortran DC
is promising~\cite{Alkan2023,Caplan2023} since it is part of the Fortran 2008 language standard and
thus can be used across CPUs and GPUs. However, it still requires modification of loops and even
mixing OpenACC to obtain maximal performance~\cite{Caplan2023}. Comparing the remaining two, recent
studies agree that OpenACC generally outperforms OpenMP Target Offloading on NVIDIA GPUs due
to compiler maturity~\cite{Deakin2020,Brunst2022}. Thus, we decide to use OpenACC for GPUs.
Although in principle OpenACC supports CPUs as well, a previous study~\cite{Deakin2020} shows that it
underperforms OpenMP by a large margin. Therefore, we use a combination of OpenACC for GPUs and
OpenMP for CPUs in this paper.

Listing~\ref{lst:offload-loop} shows the basic structure of loops after migrating to GPUs and CPUs.
As the listing shows, there are no modifications except for the OpenACC and OpenMP directives,
meeting the goal of retaining the original code structure.

On the GPU, the J and I loops are collapsed using the \verb|collapse| clause and offloaded to the
GPU. Even when the I and J loops are collapsed, they result in a total of less than $10^6$
iterations in most cases and cannot saturate the whole GPU. Furthermore, since a single kernel
invocation finishes in approximately
\qtyrange{50}{500}{\us} and kernel launches are synchronous by default in OpenACC, the kernel launch
overhead becomes a non-negligible bottleneck. To hide the launch overhead, we take advantage of the \verb|async|
clause~\cite{Chandrasekaran2017} in OpenACC. The \verb|async| clause allows the host-side thread to
continue execution, effectively hiding the kernel launch latency. In addition, the kernels are
submitted to multiple asynchronous queues in a round-robin manner to increase the utilization of the
GPU. Asynchronous queues are an abstraction of CUDA streams, and kernels submitted to different
queues are asynchronously executed from one another. This takes advantage of the fact that different
blocks can be independently updated.

The NLMNT2 routine requires additional work because it involves subroutine calls. The time
derivative of discharge fluxes in the x- and y-directions for each cell are computed in routines
XMMT and YMMT, which are invoked from the innermost loop body in NLMNT2. First, to compile these
routines for the GPU, they are marked with \verb|acc routine| directives as device functions.
Second, scalar arguments are specified with the VALUE attribute to be passed by value. This is
necessary because in Fortran, all arguments are passed by reference by default, and the compiler has
to assume loop-carried dependence. Finally, we find that enabling Link Time Optimization (LTO)
has a positive performance impact.

On the CPU, the J loop is mapped to threads and the I loop is mapped to vector lanes. To reduce
the overhead of a parallel region, we enclose the entire loop nest with a parallel region.
Furthermore, a \verb|nowait| clause is added to the \verb|omp do| directive to mitigate the
overhead of implicit synchronization. Since the compiler fails to auto-vectorize the inner
I loop in many cases, the I loop is explicitly vectorized by adding an \verb|omp simd| directive.
Similar to the GPU case, subroutine calls within the loop body become problematic since they prevent
the compiler from vectorizing the loop. Thus, an \verb|attributes forceinline|
directive\footnote{\url{https://www.intel.com/content/www/us/en/docs/fortran-compiler/developer-guide-reference/2024-1/attributes-inline-noinline-and-forceinline.html}} is added
to the XMMT and YMMT routines to force them to be inlined in the loop body.

\subsection{Communication optimization}

On the GPU, only offloading the routines for computation in the time integration loop incurs
expensive host-device memory copy for MPI communication. Thus, we utilize CUDA-aware MPI and
GPUDirect RDMA (GDR)~\cite{Potluri2013} to directly copy data between device memories and bypass
the host. To enable the use of CUDA-aware MPI and GDR, we (1) place the MPI communication buffers on
device memory, and (2) offload message packing and unpacking to the GPU.

\subsubsection{Intra-grid exchange}

Listing~\ref{lst:ptpmn-before} shows a loop in the PTP\_MN routine, which exchanges the discharge
fluxes at the boundary cells between neighboring ranks in the same grid level. This specific loop
nest packs the boundaries of two arrays VAL1 and VAL2 into a single communication buffer, BUF\_SND1.
Clearly, the loop-carried dependence on the buffer offset (ICNT) inhibits parallelization.

\begin{lstlisting}[caption={A loop in PTP\_MN before optimization.},label={lst:ptpmn-before},float]
DO J = SFTN(6), SFTN(7)
  DO I = SFTN(2), SFTN(3)
    ICNT = ICNT + 1
    BUF_SND1(ICNT) = VAL1(I,J,1)
    BUF_SND1(ICNT+JNUM) = VAL1(I,J,2)
    BUF_SND1(ICNT+2*JNUM) = VAL2(I,J)
  END DO
END DO
\end{lstlisting}

Listing~\ref{lst:ptpmn-after} shows the same loop after optimization. Since the trip counts of both
I and J loops can be pre-determined, we calculate the buffer offset from the loop indices of the two
loops. The two loops are collapsed using a \verb|collapse| clause and offloaded to the GPU as a
single kernel. A similar optimization is applied to the PTP\_Z routine as well.

\begin{lstlisting}[caption={A loop in PTP\_MN after optimization.},label={lst:ptpmn-after},float]
!$acc kernels
!$acc loop collapse(2) independent
DO J = SFTN(6), SFTN(7)
  DO I = SFTN(2), SFTN(3)
    ICNT = (I- SFTN(2) + 1) + (J - SFTN(6)) * (SFTN(3) - SFTN(2) + 1)
    BUF_SND1(ICNT) = VAL1(I,J,1)
    BUF_SND1(ICNT+JNUM) = VAL1(I,J,2)
    BUF_SND1(ICNT+2*JNUM) = VAL2(I,J)
  END DO
END DO
!$acc end kernels
\end{lstlisting}

\subsubsection{Inter-grid exchange}

Listing~\ref{lst:jnzsnd-before} shows an excerpt from the JNZSND routine. This routine sends the
water levels at the boundary cells of a child grid to its parent grid. Because each rank may have
multiple blocks and a child block can have multiple parent blocks, each rank may need to
send multiple boundary regions to multiple receiver ranks. In this routine, the outer NN1 loop iterates over the
receiver ranks, and the inner NN2 loop iterates over the boundaries of multiple blocks. The JNZ\_SND
and JNZ\_SND\_NO arrays hold information such as the receiver of the boundary and the two-dimensional range of the
boundary. The JJ loop scans the cells of a boundary and reduces the resolution by
averaging the water levels in a 3$\times$3 cell. The average water level is then written to BUFS,
the MPI communication buffer. ICNT\_WK keeps track of the current offset in the communication
buffer.

\begin{lstlisting}[caption={A loop in JNZSND before optimization.},label={lst:jnzsnd-before},float]
DO NN1 = 1, NUM_JNZ_SND_SZ ! iterate over receivers
  ICNT_WK = 0
  DO NN2 = 1, JNZ_SND_SZ(3,NN1) ! iterate over boundaries
    NPCNT = JNZ_SND_NO(NN2,NN1)
    IF(JNZ_SND(2,NPCNT)==1.OR.JNZ_SND(2,NPCNT)==2)THEN
      JSL = JNZ_SND(7,NPCNT)
      JEL = JNZ_SND(8,NPCNT)
      JJS = JNZ_SND(4,NPCNT)  - 3
      IIS = JNZ_SND(3,NPCNT)
      ...
      DO JJ=JSL,JEL
        JJS = JJS + 3
        SUMZ = 0.0
        DO J=JJS,JJS+2 ! calculate average
           DO I=IIS,IIS+2
              SUMZ = SUMZ + ZZ(I,J,2)
           END DO
        END DO
        ZZZ = SUMZ * (1.0/9.0)
        ICNT_WK = ICNT_WK + 1
        BUFS(ICNT_WK,NN1) = ZZZ
      END DO
    ELSE IF
       ...
    END IF
  END DO
END DO
\end{lstlisting}

Evidently, the incremental update of the offset ICNT\_WK introduces a loop-carried dependence and
prevents parallelization. In contrast to the intra-grid neighbor exchange routines, ICNT\_WK cannot
be computed from just the loop indices because the sizes of the boundaries are irregular. Instead,
we pre-compute a table that contains the size of each boundary region and its offset in the
communication buffer, taking advantage of the fact that the grid organization and domain
decomposition are fixed during runtime. Listing~\ref{lst:jnzsnd-after} shows the JNZSND routine
after optimization. JNZ\_BUFS\_OFS holds the pre-computed offsets. The offset of the NN2-th boundary
sent to rank NN1 is obtained by JNZ\_BUFS\_OFS(NN2,NN1). Using this pre-computed offset table, all
boundary cells can be copied to the communication buffer in parallel. We also insert
\verb|acc loop gang| and \verb|vector| directives so that a thread block is assigned to
each boundary and a thread is assigned to each cell within the boundary.

The JNZ\_RCVWAIT routine for unpacking the water depth at the receiver rank is also optimized using
a pre-computed offset table. The JNQ\_SND and JNQ\_RCVWAIT routines for exchanging the discharge
fluxes are also optimized in a similar manner.

\begin{lstlisting}[caption={A loop in JNZSND after optimization.},label={lst:jnzsnd-after},float]
DO NN1 = 1, NUM_JNZ_SND_SZ
  !$acc kernels default(present) async
  !$acc loop gang independent
  DO NN2 = 1, JNZ_SND_SZ(3,NN1)
    NPCNT = JNZ_SND_NO(NN2,NN1)
    IF(JNZ_SND(2,NPCNT)==1.OR.JNZ_SND(2,NPCNT)==2)THEN
      ...
      !$acc loop vector independent
      DO JJ=JSL,JEL
        JJS = JJS + 3
        SUMZ = 0.0
        !$acc loop seq collapse(2)
        DO J=JJS,JJS+2
          DO I=IIS,IIS+2
             SUMZ = SUMZ + ZZ(I,J,2)
          END DO
        END DO
        ZZZ = SUMZ * (1.0/9.0)
        BUFS(JNZ_BUFS_OFS(NN2,NN1)+JJ-JSL+1,NN1) = ZZZ
      END DO
    ELSE IF
      ! ...
    END IF
  END DO
  !$acc end kernels
END DO
!$acc wait
\end{lstlisting}

\subsection{Load-balancing improvement}

The original domain decomposition algorithm equalizes the total number of cells assigned to each
rank~\cite{Inoue2019}. However, since our code launches a kernel for each block,
a non-negligible cost is incurred for each block. This is apparent
in Fig.~\ref{fig:breakdown-imbalance}, which shows the runtime breakdown when executing the
simulation on 16 ranks on an NVIDIA A100 system. The breakdown clearly shows a severe load
imbalance, where on some of the ranks such as ranks 6 and 16, the NLMASS and NLMNT2 routines take a
considerably longer time than on the other ranks.

\begin{figure}
\centering
\includegraphics[scale=0.97]{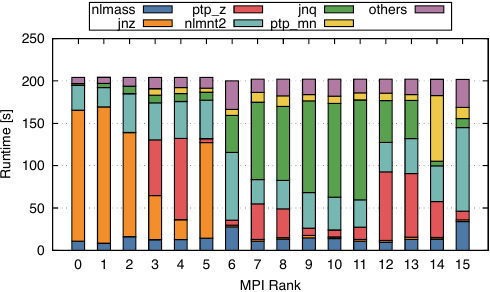}%
\caption{Breakdown of runtime before adjusting the work imbalance.}\label{fig:breakdown-imbalance}
\end{figure}

Figure~\ref{fig:decomp-before} shows the number of cells and the number of blocks assigned to each rank.
Although ranks 3 to 15 have roughly equal numbers of cells, ranks 6 and 15 have more than 16 blocks
assigned. This clearly shows that assigning too many blocks to a single rank increases its
runtime and potentially causes load imbalance. Note that ranks 0 to 2 are assigned to grid levels 1
to 3, respectively, and thus have fewer cells. This is because of the limitation of the original
code that does not allow assigning multiple grid levels to a single rank.

\begin{figure}
    \centering
    \subcaptionbox{Number of cells}
    {\includegraphics{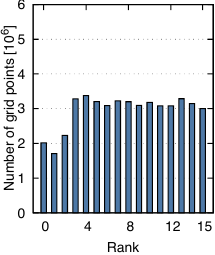}}
    \subcaptionbox{Number of blocks}
    {\includegraphics{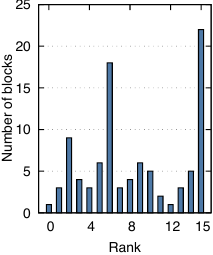}}
    \caption{Domain decomposition before optimization.}\label{fig:decomp-before}
\end{figure}

To alleviate this load imbalance, we consider two methods in this paper. In the first method, we
modify the code to launch a single kernel regardless of the number of blocks assigned to a rank. In
the second method, we quantify the per-block cost and use it to optimize the domain decomposition.
Each method is described in detail in the following sections.

\subsubsection{Merging multiple kernels}

The basic idea of this method is to collapse the outer two loops. The main reason that the
outermost loop and the inner loops cannot be trivially collapsed is that the trip counts of the
inner I and J loops depend on the KK loop. Listing~\ref{lst:outer-collapse} shows the basic loop
structure after outer loops. Our idea is to calculate the maximum trip count of the J loop and
``pad'' the iteration space in the J-dimension. In this way, the trip count of the J loop becomes
invariant, and thus the KK and J loops can be collapsed using a \verb|collapse| clause. We then
explicitly assign the collapsed KK-J loop and I loop to gangs and vectors, respectively (thread
blocks and threads in NVIDIA terminology). In the case where the outer KK loop contains multiple J-I
loops, loop fission is first applied to split the KK loop into multiple loops.

\begin{lstlisting}[caption={Basic structure of major loops after collapsing outer loops.},label={lst:outer-collapse},float]
JSZ = MAXVAL(JED_MN_X2 - JST_MN_X2)
!$acc parallel
!$acc loop collapse(2) gang
!$omp parallel do
DO KK = BLK_PROC_ST, BLK_PROC_ED
  DO J = JST_MN_X2(KK), JST_MN_X2(KK) + JSZ
    IF (J > JED_MN_X2(KK)) CYCLE

    !$acc loop vector
    !$omp simd
    DO I = IST1_OWN(KK), IED2_OWN(KK)
      ! Update cell
    END DO
    !$omp end simd
  END DO
END DO
!$omp end parallel do
!$acc end parallel loop
\end{lstlisting}

\subsubsection{Tuning the domain decomposition}

In this method, we first construct an empirical performance model of the computationally intensive
routine and use it to adjust the domain decomposition. We first implement a microbenchmark to
measure the runtime of the most time-consuming NLMNT2 routine. The benchmark repeatedly calls the
NLMNT2 routine for a given block and measures the runtime. Kernels are launched asynchronously on
multiple streams as described in Section~\ref{sec:basic-design}. Figure~\ref{fig:block-bench} plots
the runtime of NLMNT2 for a given block with respect to its number of cells. The runtime clearly
exhibits a linear trend and can be fitted by a linear function $y=1.09\cdot10^{-4}x+46.2$ with a
high $R^2$ score of 0.942.

\begin{figure}
\centering
\includegraphics[scale=0.97]{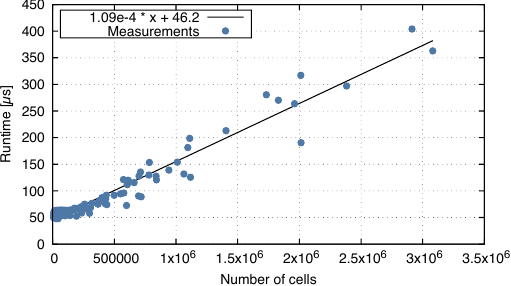}%
\caption{Runtime of NLMNT2 routine with respect to the number of cells of a block.}\label{fig:block-bench}
\end{figure}

Since each rank has one or more blocks, we estimate the runtime of a rank as the total runtime
of all blocks in a rank. We model the runtime of rank as follows:
\begin{equation}
    T = \sum_{i=1}^n{1.09\cdot10^{-4}b_i+46.2}\,[\unit{\us}],
\end{equation}
where $n$ is the total number of blocks and $b_i$ is the number of cells in the $i$-th block.
Figure~\ref{fig:perf-model} compares the actual and predicted runtime of the NLMNT2 routine using
this performance model. Although the prediction accuracy is generally high, interestingly, the actual runtime
is consistently shorter than the predicted runtime. This is likely due to a better overlap between
different blocks.

\begin{figure}
\centering
\includegraphics[scale=0.97]{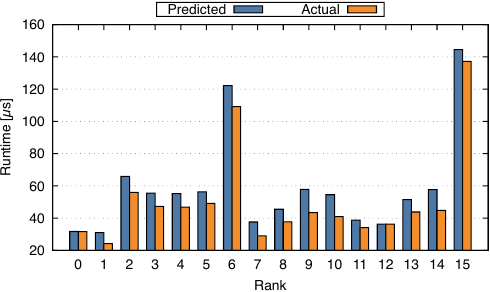}%
\caption{Runtime of NLMNT2 routine predicted with the performance model.}\label{fig:perf-model}
\end{figure}

We now use this performance model to optimize the domain decomposition such that the estimated
runtime of the NLMNT2 routine is equalized among ranks. Since each rank needs to be assigned
consecutive blocks, we assume ``separators'' between ranks as illustrated in
Fig.~\ref{fig:domain-decomp} and optimize the positions of the separators. Algorithm~\ref{alg:load-balance} shows the heuristic algorithm. The
algorithm is based on the hill climbing method, and iteratively improves the positions of the
separators, i.e., domain decomposition. In each iteration, one separator is randomly chosen and moved
to a random position between the preceding and succeeding separators. The new position is accepted if
the score function improves; otherwise, the separator is restored to its previous position. This
procedure is repeatedly executed for a fixed number of iterations.

\begin{figure}
\centering
\includegraphics[scale=0.9]{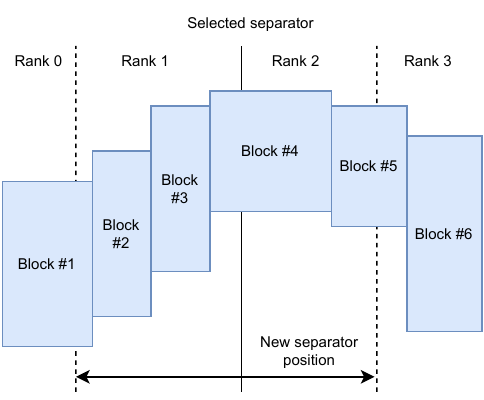}%
\caption{Overview of domain decomposition tuning.}\label{fig:domain-decomp}
\end{figure}

\begin{algorithm}[tb]
  \small
  \KwResult{List of separator positions}
  Randomly initialize positions of all separators\;
  \For{number of iterations}{
    Randomly select a separator\;
    Move the separator to a random position between the preceding and succeeding separators\;
    Compute score using the performance model\;
    \If{score does not improve}{
      Restore the position of the separator\;
    }
  }
  \caption{Algorithm for optimization the domain decomposition.}\label{alg:load-balance}
\end{algorithm}

Two score functions, which are the variance of the predicted runtime and maximum predicted runtime,
are used. This is because if the maximum runtime is used, the score changes only if the separator
adjacent to the rank with the highest predicted runtime is moved. Thus, the optimization stagnates
if only the maximum runtime is used. Contrastingly, the variance of runtime always changes if the
runtime of any rank changes. Minimizing the variance, however, does not necessarily result in
minimizing the maximum runtime. Thus, we combine the two score functions to accelerate the optimization: in the
first phase, the variance of the predicted runtime is used, and in the second phase, the maximum
predicted runtime is used.

Figure~\ref{fig:breakdown-balance} shows the runtime breakdown after optimizing the domain
decomposition. Clearly, the load balance is significantly improved, and the synchronization wait
time in the communication routines such as JNQ and PTP\_Z is reduced. As a result, the maximum
runtime of the NLMNT2 routine is reduced from \qty{99}{\second} to \qty{54}{\second}, and the total
runtime is reduced from \qty{200}{\second} down to \qty{126}{\second}. 
Figure~\ref{fig:decomp-after} shows the domain decomposition after optimization. It can be seen that
the number of cells is no longer balanced across ranks, but the maximum number of blocks is
significantly reduced. Since ranks 6 and 16 still have more blocks assigned compared to the other
ranks, the number of cells is smaller in order to offset for the larger number of blocks.

\begin{figure}
\centering
\includegraphics[scale=0.97]{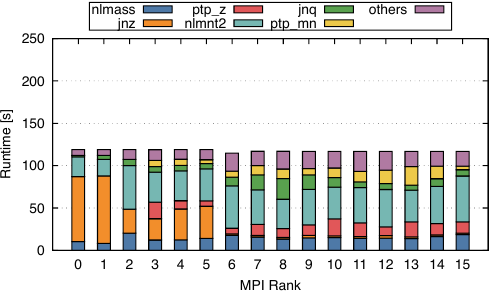}%
\caption{Breakdown of runtime after adjusting the load imbalance.}\label{fig:breakdown-balance}
\end{figure}

\begin{figure}
    \centering
    \subcaptionbox{Number of cells}
    {\includegraphics{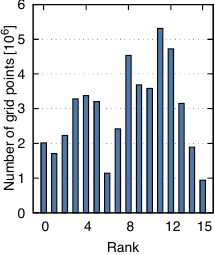}}
    \subcaptionbox{Number of blocks}
    {\includegraphics{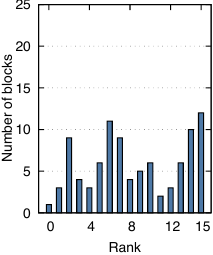}}
    \caption{Domain decomposition after optimization.}\label{fig:decomp-after}
\end{figure}

\section{Performance evaluation}\label{sec:evaluation}

In this section, we evaluate the performance of the migrated tsunami simulation code. We first
quantify the performance benefit of each optimization, and then compare the performance of our code on
multiple HPC systems using VE, CPU, and GPU.

\subsection{Evaluation environments}

In this evaluation, we use a model designed to provide tsunami inundation forecasts for Kochi
Prefecture in Japan. Kochi Prefecture is located at the Pacific coast of Japan and close to the
Nankai Trough, which is the source of Nankai megathrust earthquakes that have caused devastating
damages through history. Table~\ref{tbl:dataset} shows the organization of the grids in the
Kochi model. The model covers an area of \qty{1025}{km}$\times$\qty{1287}{km} including the
coastline of Kochi. The spatial resolution $\Delta x$ of the finest grid is \qty{10}{m} and the
$\Delta x$ of the coarsest grid is $\qty{10}{m} \cdot 3^4 = \qty{810}{m}$. The time resolution
$\Delta t$ is \qty{0.2}{s} across all grid levels.

Table~\ref{tbl:systems} summarizes the four HPC systems used in this evaluation:
AOBA-S~\cite{Takizawa2024} at Tohoku University, SQUID~\cite{Date2023} at Osaka University, and
Pegasus~\cite{Pegasus} at University of Tsukuba. AOBA-S is an SX-Aurora TSUBASA system using the
third-generation Vector Engine processors. SQUID consists of several node types. We used the CPU
nodes equipped with IceLake-generation Intel CPUs and the GPU nodes equipped with NVIDIA
A100 SXM4 GPUs. Pegasus is a supercomputer equipped with NVIDIA H100 PCIe GPUs.

\begin{table}
\centering
\caption{Grid organization of the Kochi model.}\label{tbl:dataset}
\begin{tabular}{lrrr}
\toprule
Grid level & $\Delta x$ & \# of blocks & \# of cells \\ \midrule
1          & \qty{810}{\meter} & 1            &  \num{2012940}    \\
2          & \qty{270}{\meter} & 3            &  \num{1703484}    \\
3          &  \qty{90}{\meter} & 9            &  \num{2230056}    \\
4          &  \qty{30}{\meter} & 11           &  \num{9863424}    \\
5          &  \qty{10}{\meter} & 60           & \num{31401540}    \\ \midrule
Total      &                   & 84           & \num{47211444}    \\ \bottomrule
\end{tabular}
\end{table}

\begin{table*}
\centering
\caption{HPC systems used for evaluation.}\label{tbl:systems}
\begin{tabular}{@{}lllll@{}}
\toprule
             & AOBA-S~\cite{Takizawa2024}& SQUID~\cite{Date2023} (GPU node)        & SQUID (CPU node)                   & Pegasus~\cite{Pegasus} \\ \midrule
CPU          & AMD EPYC 7763             & Intel Xeon Platinum 8368 $\times$2      & Intel Xeon Platinum 8368 $\times$2 & Intel Xeon Platinum 8468 $\times$1 \\
Memory       & DDR4 256GB                & DDR4 512GB                              & DDR4 256GB & DDR5 128GB                                                 \\
Accelerator  & NEC Vector Engine Type 30A $\times$8 & NVIDIA A100 (SXM4) $\times$8 & N/A & NVIDIA H100 (PCIe) $\times$1                                      \\
Interconnect & InfiniBand NDR200 $\times$2          & InfiniBand HDR100 $\times$4  & InfiniBand HDR200 $\times$1        & InfiniBand NDR200 $\times$1        \\
Compilers    & NEC Fortran 5.2.0         & NVIDIA HPC SDK 22.11                    & Intel oneAPI 2023.2.4 & NVIDIA HPC SDK 24.1                             \\
             &                           &                                         &                       & Intel oneAPI 2023.0.0                           \\ \bottomrule
\end{tabular}
\end{table*}

\subsection{Asynchronous and concurrent kernel launch}

To quantify the performance benefit of the asynchronous and concurrent kernel launch method employed
in our code, we vary the number of asynchronous queues and measure the performance of the NLMNT2
routine. Figure~\ref{fig:async} shows the runtime of the NLMNT2 routine normalized by its
runtime when launching the kernels synchronously. When the number of asynchronous queues is one,
i.e., the kernels are launched asynchronously but not concurrently, the speedup over synchronous
launch ranges from 1.3$\times$ to 2.0$\times$. The speedup is achieved by hiding the kernel launch
latency and primarily depends on the number of blocks assigned to the rank because it is
proportional to the number of kernel launches.

\begin{figure}
\centering
\includegraphics[scale=0.97]{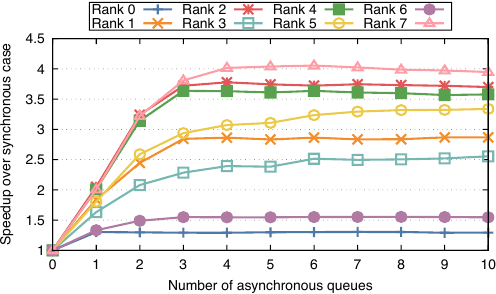}%
\caption{Runtime of NLMNT2 routine with respect to the number of asynchronous queues on NVIDIA A100.}\label{fig:async}
\end{figure}

The speedup improves with the number of asynchronous queues and saturates at four queues. The
maximum speedup ranges from 1.3$\times$ to 4.0$\times$ depending on the rank. This indicates that
the GPU is saturated when kernels are submitted concurrently to four queues.

To verify that the hardware utilization improves with the number of asynchronous queues, we measured
the GPU and memory utilization using the NVIDIA Management Library
(NVML)\footnote{\url{https://developer.nvidia.com/management-library-nvml}}. The definitions of these
metrics are as follows:
\begin{description}[\IEEEsetlabelwidth{Memory utilization}]
    \item[GPU utilization] Fraction of time where one or more kernels were running on the GPU.
    \item[Memory utilization] Fraction of time where the device memory was accessed.
\end{description}
We also tried NVIDIA Nsight Compute but found out that profiling a range including
many small and short kernel launches leads to inaccuracies.

Figure~\ref{fig:async-prof} shows the
utilization rates with respect to the number of queues. Figure~\ref{fig:async-gpu-util} shows that
the GPU is left idle when kernels are synchronously launched. The idle time is significantly
reduced when kernels are asynchronously launched. Note that the GPU utilization only reflects
whether a kernel is being executed on the GPU, and does not reflect the resource utilization of
individual kernels. Figure~\ref{fig:async-mem-util} shows that the memory utilization gradually
increases with the number of asynchronous queues, and saturates at four queues. This aligns with the
saturation of speedup observed in Fig.~\ref{fig:async}. Based on these results, we set the number of
asynchronous queues to four in the following evaluation experiments.

\begin{figure}
    \centering
    \subcaptionbox{GPU\label{fig:async-gpu-util}}{\includegraphics{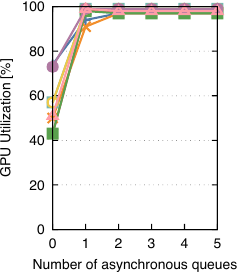}}
    \subcaptionbox{Memory\label{fig:async-mem-util}}{\includegraphics{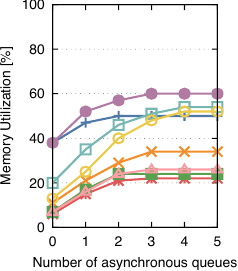}}
    \caption{GPU and memory utilization measured using NVML (legend is identical to that of Fig.~\ref{fig:async}).}\label{fig:async-prof}
\end{figure}

\subsection{Load-balancing methods}

We evaluate the effectiveness of the two methods for improving load balance by comparing the runtime
of the NLMNT2 routine before and after improving the load balance. Figure~\ref{fig:load-balance}
shows the runtime of the NLMNT2 routine on each rank. As the figure shows, both methods are able to improve the load imbalance significantly. The maximum runtime of the NLMNT2 routine in the baseline
code is \qty{139}{\us}. This is reduced down to \qty{56}{\us} when collapsing the outer loops,
and to \qty{73}{\us} when using the optimized domain decomposition. At first sight, the first
option seems to be a better choice. However, collapsing the outer loops actually degrades
performance on CPUs as shown in Fig.~\ref{fig:load-balance2}. This is because on a GPU, the
performance gain thanks to collapsing the outer loops outweighs the extra cost incurred by padding
the iteration space, but on a CPU, the load balance is already good in the baseline case and thus
collapsing the outer loop only causes a negative impact on the performance.

Of course, one could use the loop collapsed version on GPUs and the baseline version on CPUs
and VEs. However, this conflicts with the goal of having a single codebase for all platforms.
Therefore, we use the second domain decomposition method in the following evaluation.

\begin{figure}
\centering
\includegraphics[scale=0.97]{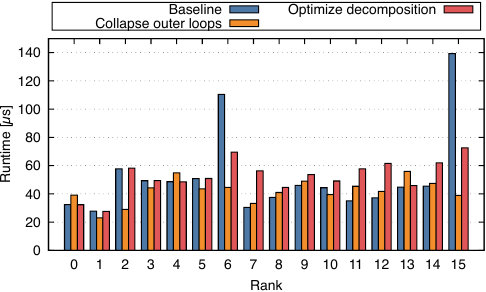}%
\caption{Runtime of NLMNT2 routine before and after improving the load balance on NVIDIA A100.}\label{fig:load-balance}
\end{figure}

\begin{figure}
\centering
\includegraphics[scale=0.97]{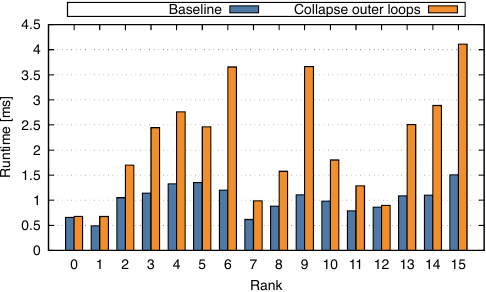}%
\caption{Runtime of NLMNT2 routine before and after improving the load balance on Intel Xeon Platinum 8468.}\label{fig:load-balance2}
\end{figure}

\subsection{Communication optimization}

To assess the impact of the intra- and inter-grid communication routine optimization, we compare a
naive implementation and the GPU-offloaded implementation of the communication routines.
Figure~\ref{fig:comm-tuning-squid} shows the runtime of a six-hour tsunami simulation using the
Kochi model on SQUID (GPU). The naive implementation copies the boundary cells between the GPU
and CPU before and after communication. Offloading the message packing and unpacking to the GPU and
utilizing GDR greatly improves the runtime by a factor of 2.96$\times$ and 1.09$\times$ on 8 and 16 ranks,
respectively. This is because the latency of host-device memory copies is eliminated and the
frequency of host-device synchronization is reduced. On 32 ranks, however, the runtime increases by
1.41$\times$ due to low inter-GPU communication performance.

After tweaking various parameters in UCX~\cite{Shamis2015}, a low-level communication library used
by Open MPI, we found that the default message size threshold for switching from eager to rendezvous
protocol was suboptimal. This issue was resolved by enabling a UCX parameter
\verb|UCX_PROTO_ENABLE|, which activates a new mechanism for automatically selecting the optimal
protocol. In addition, we restricted the InfiniBand NIC used by each GPU using the
\verb|UCX_NET_DEVICES| parameter to reduce the GPU-NIC communication latency. Since a SQUID GPU node
has four NICs and eight GPUs distributed over four PCIe switches, we configured UCX such that each
GPU uses the NIC connected to the same PCIe switch as the GPU. As a result of tuning UCX, the total
runtime improved by 1.27$\times$ on 16 ranks and 1.62$\times$ on 32 ranks.

Figure~\ref{fig:comm-tuning-pegasus} shows the simulation runtime on Pegasus (GPU). On Pegasus,
utilizing GDR  results in a speedup ranging from 2.95--3.23$\times$, larger than that on
SQUID. This is because the bottleneck routines (NLMNT2 and NLMASS) run faster on the H100 GPU than
on the A100 GPU, and thus communication tends to become a bottleneck as the number of GPUs scales
out. On Pegasus, UCX tuning is not required because \verb|UCX_PROTO_ENABLE| is enabled by default
due to a newer UCX version. In addition, GPU-NIC affinity is also not needed because each compute
node is equipped with one GPU and one NIC.

\begin{figure}
\centering
\begin{subcaptionblock}{.5\columnwidth}
\centering
\includegraphics[scale=0.97]{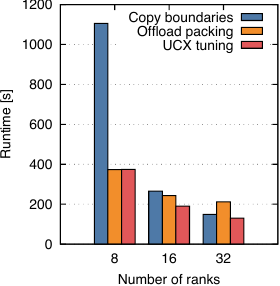}
\caption{SQUID (GPU)}\label{fig:comm-tuning-squid}
\end{subcaptionblock}%
\begin{subcaptionblock}{.5\columnwidth}
\centering
\includegraphics[scale=0.97]{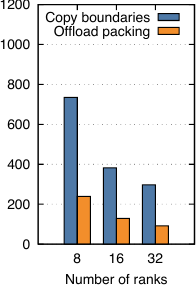}
\caption{Pegasus (GPU)}\label{fig:comm-tuning-pegasus}
\end{subcaptionblock}%
\caption{Impact of communication routine optimization on the simulation runtime.}\label{fig:comm-tuning}
\end{figure}

\subsection{Comparison across hardware platforms}

Figure~\ref{fig:system-comparison} compares the runtime for completing a six-hour tsunami simulation
using the Kochi model on SQUID, AOBA-S and Pegasus. To make a fair comparison, we fix the number of
CPUs or GPUs and compare the runtime. The MPI process count is tuned for each system, that is, one
process per socket on SQUID (CPU) and four processes per socket on Pegasus (CPU). On SQUID (GPU)
and Pegasus (GPU), one process is launched per GPU.

When using four sockets, AOBA-S marks \qty{640}{\second}, which marginally fails to achieve the
\qty{10}{\minute} deadline. SQUID (CPU) and Pegasus (CPU) achieve \qty{1636}{\second} and
\qty{1476}{\second}, respectively, both being twice as slow as AOBA-S. This is due to the excellent
memory bandwidth of the VE processor used in AOBA-S. The GPU version cannot be run because
efficiently sharing a single GPU by multiple processes requires Multi-Instance GPU (MIG) or
Multi-Process Service (MPS), both of which are unavailable on Pegasus and SQUID.

When using eight sockets, all of AOBA-S, SQUID (GPU) and Pegasus (GPU) complete within
\qty{600}{\second}, successfully meeting the 10-10-10 challenge. Pegasus (GPU) is the fastest,
followed by AOBA-S and then SQUID (GPU). This order is consistent with the order of effective memory
bandwidth of the processors used in these systems.
With 16 sockets, SQUID (CPU) and Pegasus (CPU) exhibit a super-linear speedup, and the performance
difference between the CPU-based systems and the rest becomes smaller. This is due to the large L3
cache size where the working set can fit on L3 cache as the number of sockets increases. In fact,
the L3 cache miss rate measured using the LIKWID~\cite{Treibig2010} profiling tool on 8, 16 and 32
ranks is 33\%, 14\%, 3\%, respectively. With 32 sockets, the runtime is less than \qty{3}{\minute}
on all systems, and just \qty{82}{\second} on Pegasus (GPU). At this scale, the cache hit rate is
high on all systems, making the code cache bandwidth-bound.

\begin{figure}
\centering
\includegraphics{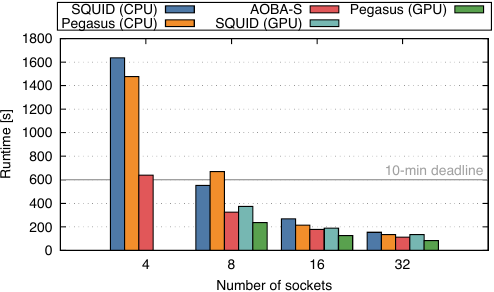}%
\caption{Runtime of a six-hour tsunami simulation using the Kochi model on different HPC
    systems.}\label{fig:system-comparison}
\end{figure}

\section{Conclusions \& Future Work}\label{sec:conclusion}

In this paper, we migrated the RTi model, a production-scale tsunami propagation and inundation simulation code
specifically designed for vector processors, to the latest CPUs and GPUs. The migration is conducted
in a minimally invasive manner, i.e., retaining the original loop structures and minimizing the
amount of code modifications, to reduce the testing and maintenance costs. A combination of
directive-based programming models and target-specific optimizations including load balance
improvement for GPUs are applied to achieve this goal successfully. We demonstrated that the
migrated code runs efficiently on recent CPUs, GPUs and vector processors. A six-hour tsunami
simulation using over 47 million cells completes in less than 2.5 minutes on 32 Intel Sapphire
Rapids CPUs and 1.5 minutes on 32 NVIDIA H100 GPUs. These results demonstrate that the code enables
broader access to accurate tsunami inundation forecasts in the future.

The load balancing method presented in this paper could be applied to other large-scale codes,
especially multi-block structured simulation codes that follow a similar parallelization strategy.
The key idea of our method is to model the kernel runtime including the GPU offloading overhead, and
use the performance model to optimize the domain decomposition with a local search. We hope to
extend this methodology to different domain decomposition schemes and computational kernels.

\section*{Acknowledgments}

This work was supported by Council for Science, Technology and Innovation (CSTI), Cross ministerial
Strategic Innovation Promotion Program (SIP), ``Development of a Resilient Smart Network System
against Natural Disasters'' Grant Number JPJ012289 (Funding agency: NIED). This work was also
supported by JSPS KAKENHI Grant Numbers JP23K11329, JP23K16890 and JP24K02945.

Part of the experiments were carried out on AOBA-S at the Cyberscience Center, Tohoku University,
SQUID at the Cybermedia Center, Osaka University, and Pegasus at the Center for Computational
Sciences, University of Tsukuba.

\bibliographystyle{IEEEtran}
\bibliography{references}

\end{document}